\documentclass[aps,twocolumn,prd,nofootinbib]{revtex4}
\usepackage{amsmath}
\usepackage{graphicx}
\usepackage{dcolumn}
\usepackage{bm}
\usepackage{amssymb}
\usepackage{latexsym}

\begin{document}

\title{Constraints on holographic dark energy from X-ray gas mass fraction of galaxy clusters}

\author{Zhe Chang$^{a}$, Feng-Quan Wu$^{b,a,c}$, and Xin
Zhang$^{b,a,c}$}

\affiliation{$^{a}$Institute of High Energy Physics, Chinese
Academy of Sciences, P.O.Box 918(4), Beijing 100049, People's
Republic of China} \affiliation{$^{b}$CCAST (World Laboratory),
P.O.Box 8730, Beijing 100080, People's Republic of China}
\affiliation{$^{c}$Graduate School, Chinese Academy of Sciences,
Beijing 100049, People's Republic of China}

\begin{abstract}
We use the {\it Chandra} measurements of the X-ray gas mass
fraction of 26 rich clusters released by Allen et al. to perform
constraints on the holographic dark energy model. The constraints
are consistent with those from other cosmological tests,
especially with the results of a joint analysis of supernovae,
cosmic microwave background, and large scale structure data. From
this test, the holographic dark energy also tends to behave as a
quintom-type dark energy.

\end{abstract}

\maketitle

Recent observations of Type Ia supernovae (SNe Ia) \cite{sn}
indicate that the expansion of the Universe is accelerating at the
present time. These results, when combined with the observations
of cosmic microwave background (CMB) \cite{wmap} and large scale
structure (LSS) \cite{sdss}, strongly suggest that the Universe is
spatially flat and dominated by an exotic component with large
negative pressure, referred to as dark energy \cite{de}. The first
year result of the Wilkinson Microwave Anisotropy Probe (WMAP)
shows that dark energy occupies about $73\%$ of the energy of our
Universe, and dark matter about $23\%$. The usual baryon matter
which can be described by our known particle theory occupies only
about $4\%$ of the total energy of the Universe. Although we can
affirm that the ultimate fate of the Universe is determined by the
feature of dark energy, the nature of dark energy as well as its
cosmological origin remain enigmatic at present. The most obvious
theoretical candidate of dark energy is the cosmological constant
$\Lambda$ which has the equation of state $w=-1$. An alternative
proposal is the dynamical dark energy (quintessence) \cite{quin}
which suggests that the energy form with negative pressure is
provided by a scalar field evolving down a proper potential. The
feature of this class of models is that the equation of state of
dark energy $w$ evolves dynamically during the expansion of the
Universe. However, as is well known, there are two difficulties
arise from all these scenarios, namely the two dark energy (or
cosmological constant) problems --- the fine-tuning problem and
the ``cosmic coincidence'' problem. The fine-tuning problem asks
why the dark energy density today is so small compared to typical
particle scales. The dark energy density is of order $10^{-47}
{\rm GeV}^4$, which appears to require the introduction of a new
mass scale 14 or so orders of magnitude smaller than the
electroweak scale. The second difficulty, the cosmic coincidence
problem, states ``Since the energy densities of dark energy and
dark matter scale so differently during the expansion of the
Universe, why are they nearly equal today''? To get this
coincidence, it appears that their ratio must be set to a
specific, infinitesimal value in the very early Universe.

Recently, considerable interest has been stimulated in explaining
the observed dark energy by the holographic dark energy model. For
an effective field theory in a box of size $L$, with UV cut-off
$\Lambda_c$ the entropy $S$ scales extensively, $S\sim
L^3\Lambda_c^3$. However, the peculiar thermodynamics of black
hole \cite{bh} has led Bekenstein to postulate that the maximum
entropy in a box of volume $L^3$ behaves nonextensively, growing
only as the area of the box, i.e. there is a so-called Bekenstein
entropy bound, $S\leq S_{BH}\equiv\pi M_p^2L^2$. This nonextensive
scaling suggests that quantum field theory breaks down in large
volume. To reconcile this breakdown with the success of local
quantum field theory in describing observed particle
phenomenology, Cohen et al. \cite{cohen} proposed a more
restrictive bound -- the energy bound. They pointed out that in
quantum field theory a short distance (UV) cut-off is related to a
long distance (IR) cut-off due to the limit set by forming a black
hole. In other words, if the quantum zero-point energy density
$\rho_X$ is relevant to a UV cut-off, the total energy of the
whole system with size $L$ should not exceed the mass of a black
hole of the same size, thus we have $L^3\rho_X\leq LM_p^2$. This
means that the maximum entropy is in order of $S_{BH}^{3/4}$. When
we take the whole Universe into account, the vacuum energy related
to this holographic principle \cite{holoprin} is viewed as dark
energy, usually dubbed holographic dark energy. The largest IR
cut-off $L$ is chosen by saturating the inequality so that we get
the holographic dark energy density
\begin{equation}
\rho_X=3c^2M_p^2L^{-2}~,\label{de}
\end{equation} where $c$ is a numerical constant, and $M_p\equiv 1/\sqrt{8\pi
G}$ is the reduced Planck mass. If we take $L$ as the size of the
current Universe, for instance the Hubble scale $H^{-1}$, then the
dark energy density will be close to the observed data. However,
Hsu \cite{hsu} pointed out that this yields a wrong equation of
state for dark energy. Li \cite{li} subsequently proposed that the
IR cut-off $L$ should be taken as the size of the future event
horizon
\begin{equation}
R_h(a)=a\int_t^\infty{dt'\over a(t')}=a\int_a^\infty{da'\over
Ha'^2}~.\label{eh}
\end{equation} Then the problem can be solved nicely and the
holographic dark energy model can thus be constructed
successfully. The holographic dark energy scenario may provide
simultaneously natural solutions to both dark energy problems as
demonstrated in Ref.\cite{li}. For related work see
\cite{holo,snfit,cmb,holoext,sfzx,snzx}.

Consider now a spatially flat FRW (Friedmann-Robertson-Walker)
Universe with matter component $\rho_m$ (including both baryon
matter and cold dark matter) and holographic dark energy component
$\rho_X$, the Friedmann equation reads
\begin{equation}
3M_p^2H^2=\rho_m+\rho_{X}~,
\end{equation} or equivalently,
\begin{equation}
{H^2\over H_0^2}=\Omega_m^0a^{-3}+\Omega_X{H^2\over
H_0^2}~.\label{Feq}
\end{equation}
Note that we always assume spatial flatness throughout this paper
as motivated by inflation. Combining the definition of the
holographic dark energy (\ref{de}) and the definition of the
future event horizon (\ref{eh}), we derive
\begin{equation}
\int_a^\infty{d\ln a'\over Ha'}={c\over
Ha\sqrt{\Omega_X}}~.\label{rh}
\end{equation} We notice that the Friedmann
equation (\ref{Feq}) implies
\begin{equation}
{1\over Ha}=\sqrt{a(1-\Omega_X)}{1\over
H_0\sqrt{\Omega_m^0}}~.\label{fri}
\end{equation} Substituting (\ref{fri}) into (\ref{rh}), one
obtains the following equation
\begin{equation}
\int_x^\infty e^{x'/2}\sqrt{1-\Omega_X}dx'=c
e^{x/2}\sqrt{{1\over\Omega_X}-1}~,
\end{equation} where $x=\ln a$. Then taking derivative with respect to $x$ in both
sides of the above relation, we get easily the dynamics satisfied
by the dark energy, i.e. the differential equation about the
fractional density of dark energy,
\begin{equation}
{d\Omega_X\over d\ln a}=\Omega_X(1-\Omega_X)(1+{2\over
c}\sqrt{\Omega_X})~. \label{deq}\end{equation} This equation
describes behavior of the holographic dark energy completely, and
it can be solved exactly \cite{li,snfit}. From the energy
conservation equation of the dark energy, the equation of state of
the dark energy can be given \cite{li}
\begin{equation}
w=-1-{1\over 3}{d\ln\rho_X\over d\ln a}=-{1\over 3}(1+{2\over
c}\sqrt{\Omega_X})~.\label{w}
\end{equation} Note that the formula
$\rho_X={\Omega_X\over 1-\Omega_X}\rho_m^0a^{-3}$ and the
differential equation of $\Omega_X$ (\ref{deq}) are used in the
second equal sign. It can be seen clearly that the equation of
state of the holographic dark energy evolves dynamically and
satisfies $-(1+2/c)/3\leq w\leq -1/3$ due to $0\leq\Omega_X\leq
1$. In this sense, this model should be attributed to the class of
dynamical dark energy models even though without quintessence
scalar field. The parameter $c$ plays a significant role in this
model. If one takes $c=1$, the behavior of the holographic dark
energy will be more and more like a cosmological constant with the
expansion of the Universe, and the ultimate fate of the Universe
will be entering the de Sitter phase in the far future. As is
shown in Ref.\cite{li}, if one puts the parameter
$\Omega_X^0=0.73$ into (\ref{w}), then a definite prediction of
this model, $w_0=-0.903$, will be given. On the other hand, if
$c<1$, the holographic dark energy will behave like a quintom-type
dark energy proposed recently in Ref.\cite{quintom}, the amazing
feature of which is that the equation of state of dark energy
component $w$ crosses the phantom divide, $-1$, i.e. it is larger
than $-1$ in the past while less than $-1$ near today. The recent
fits to current SNe Ia data with parametrization of the equation
of state of dark energy find that the quintom-type dark energy is
mildly favored \cite{starob,huterer}. Usually the quintom dark
energy model is realized in terms of double scalar fields, one is
a normal scalar field and the other is a phantom-type scalar field
\cite{quintom1} (for quintom model see e.g. \cite{quintom2}).
However, the holographic dark energy in the case $c<1$ provides us
with a more natural realization for the quintom picture. If $c>1$,
the equation of state of dark energy will be always larger than
$-1$ such that the Universe avoids entering the de Sitter phase
and the Big Rip phase. Hence, we see explicitly, the determination
of the value of $c$ is a key point to the feature of the
holographic dark energy as well as the ultimate fate of the
Universe.

The holographic dark energy model has been tested and constrained
by various astronomical observations \cite{snfit,cmb,snzx}. In a
recent work \cite{snzx}, it has been explicitly shown that
regarding the latest supernova data as well as the CMB and LSS
data, the holographic dark energy behaves like a quintom-type dark
energy. This indicates that the numerical parameter $c$ in the
model is less than 1. The best fit results provided by \cite{snzx}
are: $c=0.81$, $\Omega_m^0=0.28$, and $h=0.65$, which lead to the
present equation of state of dark energy $w_0=-1.03$ and the
deceleration/acceleration transition redshift $z_T=0.63$. It is
necessary to test dark energy models and constrain their
parameters using as many techniques as possible. Different tests
might provide different constraints on the parameters of the
model, and a comparison of results determined from different
methods allows us to make consistency checks. In this Letter, we
use the X-ray gas mass fraction of rich clusters, as a function of
redshift, to constrain the holographic dark energy model, and to
compare the results with the previous analysis.

The matter content of the largest clusters of galaxies is thought
to provide an almost fair sample of the matter content of the
Universe. A comparison of the gas mass fraction of galaxy
clusters, $f_{\rm gas}=M_{\rm gas}/M_{\rm tot}$, inferred from
X-ray observations, with $\Omega_b^0$ determined by
nucleosynthesis can be used to constrain the density parameter of
the Universe $\Omega_m^0$ directly \cite{measurem}. Sasaki
\cite{sasaki} and Pen \cite{pen} were the first to describe how
the $f_{\rm gas}$ data of clusters of galaxies at different
redshifts could also, in principle, be used to constrain the
geometry and, therefore, dark energy relevant parameters of the
Universe. The geometrical constraint arises from the fact that the
measured $f_{\rm gas}$ values for each galaxy cluster depend on
the assumed angular diameter distances to the clusters as $f_{\rm
gas}\propto d_A^{3/2}$. The measured $f_{\rm gas}$ values should
be invariant with redshift \cite{sasaki,pen,eke} when the
reference cosmology used in making the measurements matches the
true, underlying cosmology. The first successful application of
such a test to constrain cosmological parameters was carried out
by Allen et al. \cite{A02}; see also \cite{A03,A04,lima,chen,zhu}
and references herein. Note that the optically luminous galaxy
(stellar) mass in clusters is about $0.19\sqrt{h}$ times the X-ray
emitting gas mass, thus $\Omega_b^0=\Omega_m^0 f_{\rm
gas}(1+0.19\sqrt{h})$. In what follows we use the $f_{\rm gas}$
values, determined by Allen et al. \cite{A04} from {\it Chandra}
observational data, to constrain the parameters of the holographic
dark energy model. The redshifts of the 26 clusters range from
0.08 to 0.89.

Following \cite{A02,A03,A04,lima,chen,zhu}, we fit the $f_{\rm
gas}$ data to the holographic dark energy model described by
\begin{equation}
f_{\rm gas}^{\rm mod}(z)={b\Omega_b^0\over
(1+0.19\sqrt{h})\Omega_m^0}\left[{h\over 0.5}{d_A^{\rm
SCDM}(z)\over d_A^{\rm
mod}(z;\Omega_m^0,c)}\right]^{3/2}~\label{fgas},
\end{equation}
where $d_A^{\rm mod}$ and $d_A^{\rm SCDM}$ are the angular
diameter distances to the clusters in the current holographic
model and reference SCDM cosmology, respectively, and $b$ is a
bias factor motivated by gasdynamical simulations which suggest
that the baryon fraction in clusters is slightly lower than for
the Universe as a whole(see \cite{A03,A04} and references herein
for detailed discussions). The angular diameter distances to the
clusters are defined as
\begin{equation}
d_A=H_0^{-1}(1+z)^{-1}\int_0^z{dz'\over E(z')}~,
\end{equation}
where $H_0^{-1}$ (here we use the natural unit, namely the speed
of light is defined to be 1) represents the Hubble distance with
value $H_0^{-1}=2997.9h^{-1}$ Mpc, and $E(z)=H(z)/H_0$ can be
obtained from (\ref{Feq}), expressed as
\begin{equation}
E(z)=\left({\Omega_m^0(1+z)^3\over 1-\Omega_X}\right)^{1/2}~.
\end{equation} Note that for the holographic dark energy model
the dynamical behavior of $\Omega_X$ is determined by (\ref{deq});
while for the SCDM model we have $\Omega_X=0$ and $\Omega_m^0=1$.
It should be pointed out that the $f_{\rm gas}$ data used here are
determined assuming an SCDM model with $h=0.5$. Hence there
appears an $h/0.5$ factor in (\ref{fgas}). We use the same
Gaussian priors in our computation as \cite{A04,chen} with
$h=0.72\pm 0.08$, $\Omega_b^0 h^2=0.0214\pm 0.002$, and
$b=0.824\pm 0.089$, all $1\sigma$ errors.

To constrain the parameters of the holographic dark energy model,
we use a $\chi^2$ statistic
\begin{eqnarray}
\chi^2&= &\sum_{i=1}^{26}{[f_{\rm gas}^{\rm mod}(z_i;P)-f_{{\rm
gas},i}]^2\over\sigma_{f_{\rm gas},i}^2}+
\left({\Omega_b^0h^2-0.0214\over
0.002}\right)^2\nonumber\\&+&\left({h-0.72\over
0.08}\right)^2+\left({b-0.824\over 0.089}\right)^2~,
\end{eqnarray}
where $f_{\rm gas}^{\rm mod}(z_i;P)$ is computed by the
holographic dark energy model using (\ref{fgas}), and $f_{{\rm
gas},i}$ and $\sigma_{f_{\rm gas},i}$ are the measured value and
error from \cite{A04} for a cluster at redshift $z_i$,
respectively. The computation of $\chi^2$ is carried out in a
five-dimensional space, for the five parameters $P=(\Omega_m^0, c,
h, \Omega_b^0h^2, b)$. The probability distribution function
(likelihood) of $\Omega_m^0$ and $c$ is determined by
marginalizing over the ``nuisance''parameters
\begin{equation}
{\cal L}(\Omega_m^0,c)=\int dhd(\Omega_b^0h^2)db~ e^{-\chi^2/2}~,
\end{equation}
where the integral is over a large enough range of $h$,
$\Omega_b^0h^2$, and $b$ to include almost all the probability. We
now compute ${\cal L}(\Omega_m^0,c)$ on a two-dimensional grid
spanned by $\Omega_m^0$ and $c$. The $68.3\%$, $95.4\%$, and
$99.7\%$ (namely 1, 2, and 3 $\sigma$) confidence contours consist
of points where the likelihood equals $e^{-2.31/2}$,
$e^{-6.18/2}$, and $e^{-11.83/2}$ of the maximum value of the
likelihood, respectively.

\begin{figure}[h]
\begin{center}
\includegraphics[scale=0.8]{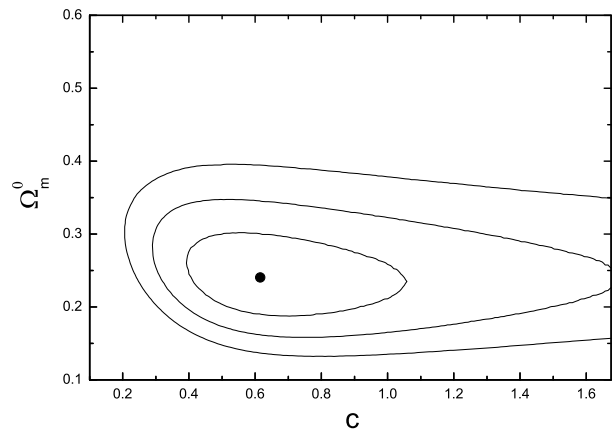}
\caption[]{Confidence level contours of $68.3\%,~95.4\%$ and
$99.7\%$ in the $(c,\Omega_m^0)$ plane. The 1 $\sigma$ fit values
for the parameters are: $\Omega_m^0=0.24^{+0.06}_{-0.05}$ and
$c=0.61^{+0.45}_{-0.21}$, and the minimum value of $\chi^2$ in the
five-dimensional parameter space is: $\chi_{\rm min}^2=25.00$. }
\end{center}
\end{figure}

Figure 1 shows our main results. We plot $68.3\%$, $95.4\%$, and
$99.7\%$ confidence level contours in the $(c, \Omega_m^0)$ plane.
The best fit happens at $c=0.61$, $\Omega_m^0=0.24$, $h=0.73$,
$\Omega_b^0 h^2=0.0212$, and $b=0.812$, with $\chi_{\rm
min}^2=25.00$. These results are in accordance with those obtained
in \cite{chen} where some common results, $\Omega_m^0=0.24$ and
$\chi^2_{\rm min}\sim 25$, were got from $f_{\rm gas}$ fits to
three models --- $\Lambda$CDM model, XCDM parametrization, and
$\phi$CDM model (quintessence with power law potential). From
Figure 1 we see clearly that the quality of the $f_{\rm gas}$
constraints is much better than that of the SNe Ia constraints
(see Figure 2 of \cite{snzx}), namely the contours are tighter
than those derived from SNe Ia data. We find, however, that the
$f_{\rm gas}$ constraints on the holographic model are consistent
with those from a joint analysis of SNe Ia, CMB, and LSS data, but
the constraints from the latter are tighter; see Figure 6 of
\cite{snzx} for comparison. The 1 $\sigma$ fit values for $c$ and
$\Omega_m^0$ are: $c=0.61^{+0.45}_{-0.21}$ and
$\Omega_m^0=0.24^{+0.06}_{-0.05}$. We notice that the fit value of
$c$ is less than 1 in 1 $\sigma$ range, though it can be slightly
larger than 1. This implies that according to the $f_{\rm gas}$
constraints the holographic dark energy basically behaves as a
quintom-type dark energy in 1 $\sigma$ range.

\begin{figure}[h]
\begin{center}
\includegraphics[scale=0.8]{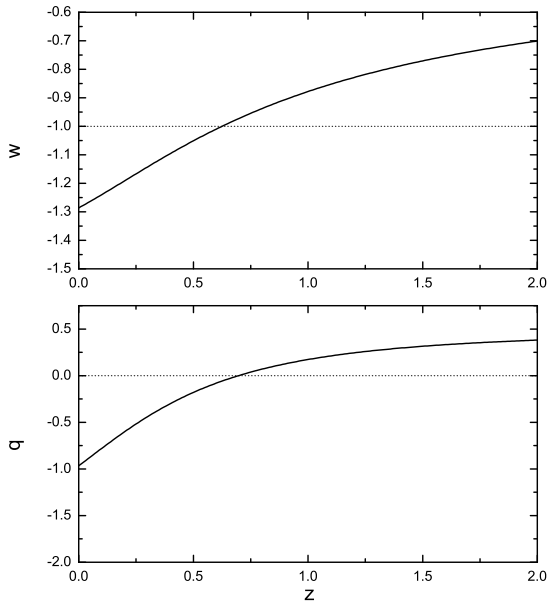}
\caption[]{Equation of state of dark energy $w$ and deceleration
parameter $q$, versus red-shift $z$, from the best fit of the
$f_{\rm gas}$ test. }
\end{center}
\end{figure}

\begin{figure}[h]
\begin{center}
\includegraphics[scale=0.8]{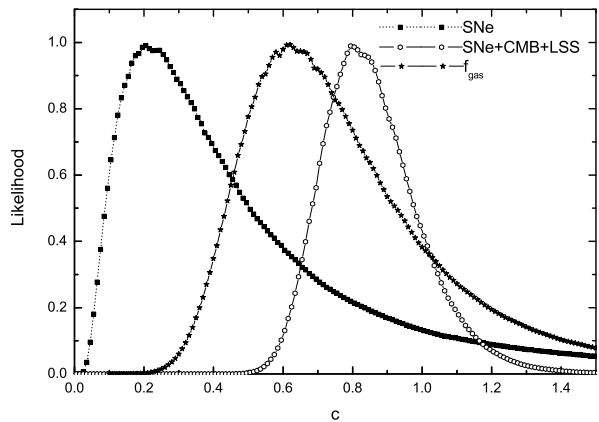}
\caption[]{Likelihood distributions of parameter $c$ in the fits
of SNe only, SNe+CMB+LSS, and $f_{\rm gas}$ data. }
\end{center}
\end{figure}

We now discuss about the cosmological consequences led by the best
fit results of the $f_{\rm gas}$ data analysis. The evolutions of
the equation of state of dark energy and the deceleration
parameter of the Universe corresponding to the best fit are shown
in Figure 2. From this figure, we see that the equation of state
of dark energy $w$ has a value of $w_0=-1.29$ and the deceleration
parameter $q$ has a value of $q_0=-0.97$ at present. The typical
characteristic of the quintom-type dark energy is that the
equation of state can cross $-1$. For this case, the crossing
behavior $(w(z_C)=-1)$ occurs at a redshift of $z_C=0.62$. In
addition, the transition from deceleration to acceleration
$(q(z_T)=0)$ occurs at the redshift $z_T=0.70$. Comparing our
plots in Figure 2 with the model-independent plots in
\cite{starob}, we find that the holographic plots for the $c=0.61$
case are in good agreement with those model-independent plots for
the redshift range $z=0-2$. On the whole, the results derived from
the $f_{\rm gas}$ constraints are consistent with those from other
cosmological tests. The parameter $c$ which plays an important
role in the holographic dark energy model is demonstrated to be
less than 1 basically in 1 $\sigma$ range, which shows that the
holographic dark energy tends to behave as quintom-type dark
energy in the cosmological evolution. For comparing the
probability distribution of the parameter $c$ determined by
different cosmological tests, we show in Figure 3 the likelihood
plots of $c$ corresponding to constraints from SNe, SNe+CMB+LSS (
for detail see \cite{snzx}), and $f_{\rm gas}$ data, respectively,
by furthermore marginalizing over the ``nuisance'' parameter
$\Omega_m^0$. We see that the X-ray data provide a fairly good way
for constraining the holographic dark energy.

In summary, we used in this Letter the recent X-ray cluster gas
mass fraction data from the {\it Chandra X-Ray Observatory} to
constrain the parameters of the holographic dark energy model. We
considered a spatially flat FRW Universe with matter and
holographic dark energy. For the holographic dark energy model,
the numerical parameter $c$ plays a very important role in
determining the evolutionary behavior of the space-time as well as
the ultimate fate of the Universe. The constraints from the
$f_{\rm gas}$ data show that in 1 $\sigma$ range the parameter $c$
is basically less than 1, which implies that the holographic dark
energy tends to behave as a quintom-type dark energy. These
constraints are consistent with those derived from other
cosmological tests. The $f_{\rm gas}$ data are proven to be
efficacious in constraining dark energy. We hope that the future
$f_{\rm gas}$ data should provide an even tighter constraint on
holographic dark energy model and other dark energy models.

\begin{acknowledgments}
One of us (X.Z.) is grateful to Ling-Mei Cheng for useful
discussions. This work was supported by the Natural Science
Foundation of China (Grant No. 111105035001).
\end{acknowledgments}



\end{document}